\documentstyle[psfig]{l-aa}
\def\kmss{km~s$^{-1}$ }
\def\kms{km~s$^{-1}$}
\def\disp{\ifmmode {{\langle}v^2{\rangle}^{1/2}}
           \else {${\langle}v^2{\rangle}^{1/2}$} \fi}
\def\dispr{\ifmmode {{\langle}v^2_R{\rangle}^{1/2}}
           \else {${\langle}v^2_R{\rangle}^{1/2}$} \fi}
\def\disprn{\ifmmode {{\langle}v^2_R{\rangle}^{1/2}_{R=0}}
           \else {${\langle}v^2_R{\rangle}^{1/2}_{R=0}$} \fi}
\def\disprh{\ifmmode {{\langle}v^2_R{\rangle}^{1/2}_{R=h}}
           \else {${\langle}v^2_R{\rangle}^{1/2}_{R=h}$} \fi}
\def\dispt{\ifmmode {{\langle}v^2_{\Theta}{\rangle}^{1/2}}
           \else {${\langle}v^2_{\Theta}{\rangle}^{1/2}$} \fi}
\def\dispz{\ifmmode {{\langle}v^2_z{\rangle}^{1/2}}
           \else {${\langle}v^2_z{\rangle}^{1/2}$} \fi}
\def\dispzn{\ifmmode {{\langle}v^2_z{\rangle}^{1/2}_{R=0}}
           \else {${\langle}v^2_z{\rangle}^{1/2}_{R=0}$} \fi}
\def\vmaxd{\ifmmode {v^{\rm disc}_{\rm max}}
           \else {$v^{\rm disc}_{\rm max}$} \fi}
\def\vmaxds{\ifmmode {v^{\rm disc}_{\rm max}}
           \else {$v^{\rm disc}_{\rm max}$ } \fi}
\def\ct{ {\langle}ct{\rangle} }
\def\hi{H\,{\sc i} }

\begin{document}
\thesaurus{11(11.07.1 11.11.1 11.12.2 11.19.2)}
\title{The maximum rotation of a galactic disc}
\author{Roelof Bottema}
\institute{Kapteyn Astronomical Institute,
P.O. Box 800,
NL-9700 AV Groningen,
The Netherlands}
\date{Received date1; accepted date2}
\maketitle
\begin{abstract}
The observed stellar velocity dispersions of galactic discs show that
the maximum rotation of a disc is on average 63\% of the observed
maximum rotation. This criterion can, however, not be applied to small
or low surface brightness (LSB) galaxies because such systems show,
in general, a continuously rising rotation curve until the
outermost measured radial position. That is why a general
relation has been derived, giving the maximum rotation for a disc
depending on the luminosity, surface brightness, and colour of the disc.
As a physical basis of this relation serves
an adopted fixed mass-to-light ratio as a function of colour.
That functionality is consistent with results from population
synthesis models and its absolute value is determined from the
observed stellar velocity dispersions. The derived maximum
disc rotation  is compared with a number of observed maximum
rotations, clearly demonstrating the need for appreciable amounts
of dark matter in the disc region and even more so for LSB galaxies.
Matters have been illustrated for two examples; the galaxy NGC 6503
and LSB galaxy NGC 1560.
\end{abstract}
\section{Introduction}
Rotation curves derived from neutral hydrogen observations 
at the outer regions of spiral
galaxies unambiguously show that substantial amounts of
dark matter are required (Bosma 1978; Begeman 1987, 1989).
Any physically reasonable distribution of this dark matter
necessitates the presence of at least some of that in the 
inner optical disc region, contributing in some degree to the
total rotation in that region.
Unfortunately, from the observed rotation curve and light
distribution one cannot a priori determine the
ratio of dark to luminous matter (van Albada et al. 1985).
This means that the 
M/L ratio of the disc cannot be determined from a rotation curve analysis
only. There are arguments which might lead to the so called
``maximum disc hypothesis'' (hereafter: md hypothesis; 
van Albada \& Sancisi 1986; Freeman 1992),
favouring a maximum possible rotational
contribution of the disc. However, no hard evidence exists
proving this hypothesis. To determine not only the amount
of dark matter in a galaxy, but also to constrain mass models of
dark matter and galaxy formation scenarios (Katz \& Gunn 1991),
it is of the utmost importance to know the relative rotational
contribution of the disc.

Stellar velocity dispersions provide a direct measure
of the local surface density of a disc and from that the
disc rotation can be calculated.
For a sample of 12 disc-dominated galaxies such
dispersions have been measured. Results are
summarized, discussed and analysed by Bottema (1993, hereafter B93).
It appears that the magnitude of the stellar velocity dispersions,
both in the radial and vertical direction, is proportional to
the square root of the surface density. That can be explained when for 
a stellar disc, a constant M/L ratio is combined with the observed
constancy of the scaleheight as a function of radius (van der Kruit \&
Searle 1981a,b, 1982). To compare the stellar kinematics of different
galactic discs the dispersion was parameterized by fitting a radial 
relation to the observations and taking the dispersion value at, 
for instance, one scalelength. Comparison of the inclined and face-on
systems showed that the ratio of vertical to radial dispersion is close
to 0.6, as is observed in the solar neighbourhood. Moreover, larger and
more massive discs have larger velocity dispersions. These matters are
illustrated in Fig. 1, for the sample of 12 galaxies with
observed dispersions.

For an exponential disc a simple relation can be derived
(Freeman 1970; B93) between the maximum rotation of a disc and the
vertical velocity dispersion $\dispz$:
\begin{equation}
\vmaxd = 0.88\; \dispzn \sqrt{ {{h}\over{z_0}} }, 
\end{equation}
where the maximum is reached at a radius of 2.2$h$. This relation 
involves only the radial scalelength ($h$) and the vertical sech$^2$
scale parameter $z_0$ (van der Kruit \& Searle 1981a; see also Eq. 26),
which is approximately equal to twice the exponential scaleheight for
suitable distances above the plane. The observed dispersions
then allow the calculation of the rotation which the disc can supply
to the total galactic rotation. For a reasonable $h/z_0$ behaviour
it appears that the maximum disc contribution is 
63\% $\pm$ 10\%, roughly independent of the mass of a galaxy. The
missing rotation then has to be supplied by the dark halo (and the 
bulge, if present). Note that this 63\% criterion does \underbar{not}
depend on the colour and surface brightness of the disc. When
applying the md hypothesis the disc contributes at 2.2 scalelengths 
between 85 to 90\% of the observed maximum rotation. The value following
from the observed velocity dispersion is considerably lower, which
leads to disc masses and M/L ratios being a factor of two smaller.
Nevertheless, with the 63\% contribution, the disc is still dominant
in the inner regions.

Ideally if one would like to know the disc rotational contribution
one should take a spectrum of the galaxy and determine the
velocity dispersion of the disc. That is, unfortunately, quite difficult
and time consuming. How then to determine quickly the amount of disc
rotation and disc mass? For a ``normal'' galaxy or sample with normal
galaxies the 63\% criterion can safely be assumed although with
an error of 10\%. Here normal means
comparable to the sample in B93 with $v^{\rm obs}_{\rm max}$
$\ga$ 100 \kms, B-V around 0.7 and central surface brightness (${\mu}_0$)
close to the value of Freeman's (1970) law; 
${\mu}_0 = {\mu}_{0,F}$ = 21.65 B-mag. arcsec$^{-2}$.
Problems arise for galaxies which are small, faint, or blue.
Small galaxies are often faint, meaning having a low surface
brightness (LSB) although occasionally also large LSB galaxies are found
(de Blok \& McGaugh 1996). Such small and/or faint galaxies generally
have rotation curves which do not reach a maximum velocity
over the observed radial extent (Casertano \& van Gorkom 1991).
Obviously the 63\% criterion then has no meaning, but could the
description be generalized in some way?
\begin{figure}[htbp]
\psfig{figure=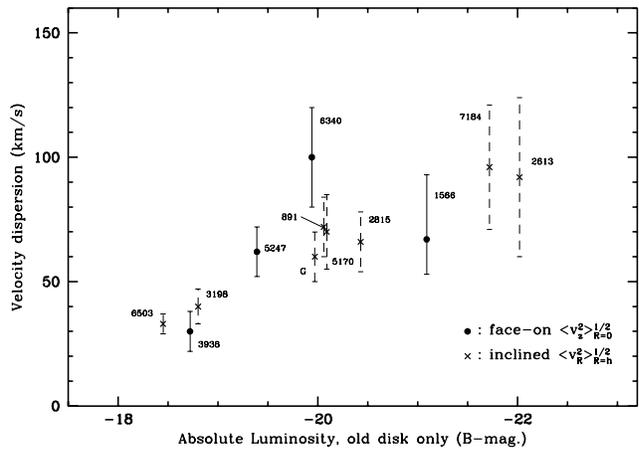,width=8.8cm}
\caption{
Observed disc stellar velocity dispersions of a sample of 12
spiral galaxies (B93) as a function of the absolute luminosity of
the old disc population (see Sect. 2). 
The disc dispersion is parameterized in
such a way that for $\dispz / \dispr = 0.6$ data for
face-on and inclined systems should fall on the same relation; as
appears to be the case. Obviously, the brighter and more 
massive galaxies have larger velocity dispersions.
}
\end{figure}

Most of the light of very blue galaxies originates from a
young stellar population which has a negligible 
mass and consequently this light
is not representative for the massive stellar population
which determines the velocity dispersion. Therefore a population
correction is needed when comparing velocity dispersion with the
brightness of a galaxy. This was dealt with in B93 by assigning
a so-called ``old disc population'' absolute magnitude ($M_{\rm od}$)
to a disc. An extensive description of this procedure is given in Sect. 2
such that for a galaxy with arbitrary colour the disc rotational 
contribution can be determined. 

The remainder of this paper deals with the construction of a general
description for the maximum rotation of a galactic disc. To that aim in
Sect. 2 a simple but adequate population correction procedure is discussed.
Section 3 analyses the situation for a single colour, fixed central
surface brightness
disc and in Sect. 4 this is generalized for an arbitrary disc.
The inferred mass-to-light ratio is calculated in Sect. 5,
and Sect. 6 describes two examples: the normal spiral NGC 6503
and LSB galaxy NGC 1560. Finally in Sect. 7 the method and its 
applicability is discussed and conclusions are formulated.
Throughout a Hubble constant of 75 \kmss Mpc$^{-1}$ is adopted. 
\section{The old disc population}
Velocity dispersions measure the local mass density in the disc.
To make any comparison between dynamical quantities derived from the 
dispersion and the emitted light one would like a reasonable 
indication of the light emitted by the population
that contains nearly all the mass in the disc.
In principle, the light of any young, massless population should
be subtracted.
A mass-to-light ratio $(M/L)_{\rm od}$ can then be assigned
to the remaining old disc population
and the 
principal assumption for the remainder of this paper will be
that this $(M/L)_{\rm od}$ is the same for all galactic discs.
This is equivalent to assuming an equal M/L ratio for galaxies having
the same colour. Such an assumption is perfectly reasonable
and several arguments for this are given in B93. The basic underlying
hypothesis is that for the low mass stars the IMF is the same for all discs
and that the range of metallicities is not too broad. Although there is
no proof that the low mass end of the IMF is universal, there is 
certainly no indication for the contrary (Laughlin \& Bodenheimer 1993;
Wyse 1995).
To obtain the luminosity of the old disc, the bulge light, of course,
also has to be subtracted from the total light of the galaxy.

In Bottema (1988) a so called ``poor man's'' population synthesis (pmps)
was performed to treat the problem of colour gradients in the disc
of NGC 3198. This pmps has been applied to the galaxy sample in B93
and will presently be described and discussed in detail. 
A galactic disc is assumed to consist of only two stellar populations;
an old disc population and a young disc population defined as:

\medskip
The old disc population:
\begin{itemize}
\item contains \underbar{all} the mass in a disc.
\item has B-V = 0.97
\end{itemize}

\medskip
The young disc population:
\begin{itemize}
\item contains no mass.
\item has B-V = -0.03.
\end{itemize}
Using the observed B-V colour the amount of light from each component
can be determined in the B or the V band. This is illustrated in Fig. 2,
where as a function of B-V colour the ratio of old disc to total disc
light is presented. For example, for B-V = 0.6, 52\% of the light
in the B-band originates from the old disc population. The absolute 
magnitude of the old disc population in the B-band ($M^B_{\rm od}$) is
related to the total magnitude in B ($M^B_{\rm tot}$) as
\begin{equation}
M^B_{\rm od} = M^B_{\rm tot} - ct,  
\end{equation}
with the correction term $ct$ given by
\begin{equation}
ct = 2.5\; {\rm log}_{10}\left[ {{ 1 - 0.973 W}\over{1.470W}} \right],
\end{equation}
and
\begin{equation}
W = 10^{-0.4(B-V)}. 
\end{equation}
This treatment has its shortcomings
A preferable complete population synthesis, however, is
much more complicated (Larson \& Tinsley 1978; Bruzual \& Charlot 1993;
Worthey 1994),
both in handling such models and in applying it to the present
specific situation. Even then effects of dust and metallicity
are not or only partially included. In Sect. 7 the influence of
particularly these two parameters on the employed pmps is investigated.
It appears that a good assessment of the effects can be made;
which counteract one another and are individually always
below a 15\% level for the galaxies of interest.
\begin{figure}[htbp]
\psfig{figure=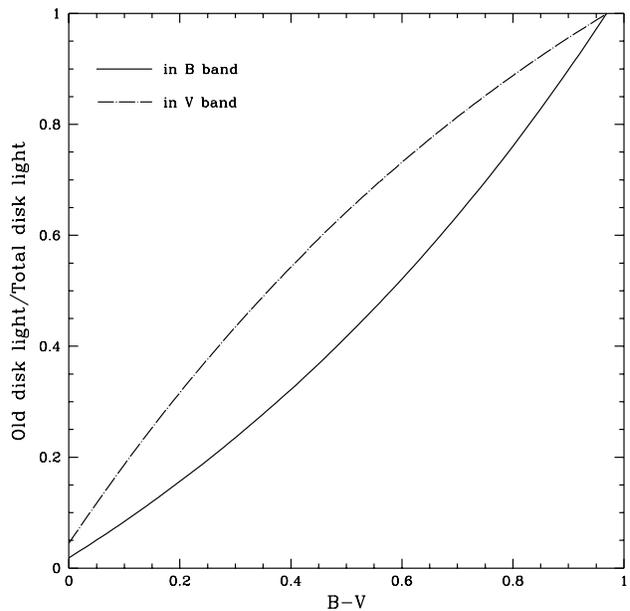,width=8.8cm}
\caption{
The proportion of the light of the old stellar, mass containing,
population for different B-V colours.
}
\end{figure}

The pmps gives a total mass-to-light ratio $(M/L)_B$ proportional
to $10^{0.4ct}$, and after stellar velocity dispersions are compared with
the luminosity of galaxies in the following sections, the absolute scale
of the mass-to-light ratio is fixed at
\begin{equation}
(M/L)_B = 2.84\; 10^{0.4ct} = 1.93\; 10^{0.4(B-V)} - 1.88,
\end{equation}
in solar units.
This can be compared with predictions of stellar population models.
Unfortunately such models cannot predict the absolute scale of 
the mass-to-light ratio, only the functionality with colour. This is caused
by the uncertainty of the IMF at the low mass end. For instance,
$M/L \propto m_l^{1-x}$, proportional to the low mass cutoff $(m_l)$
to the power $1-x$, where $x = 1.35$ for a Salpeter (1955) IMF.
This means that mass-to-light ratios can be increased simply by adding
more low mass stars. The uncertainty in absolute $M/L$ ratio is even
more increased because $M/L$ ratios derived from observations are 
proportional to the adopted Hubble constant. Hence $M/L$ ratios can only
be compared differentially and presently values will be fixed at
$(M/L)_B = 1.79$ for B-V = 0.7 as given by Eq. (5). Results of the pmps
are compared with population synthesis models of Tinsley (1981, hereafter
T81) and those of Jablonka \& Arimoto (1992, hereafter JA92). Presented
in Fig. 3 is a comparison of the B-band mass-to-light ratio as a function
of B-V, for the pmps, T81 and JA92. In all cases $(M/L)_B$ is increasing
towards redder colours, though for the T81 models more steeply
than the others. It is striking that, despite its simplicity, the
pmps shows a nearly identical trend as the other more sophisticated models.
This supports the applicability of the method.
\begin{figure}[htbp]
\psfig{figure=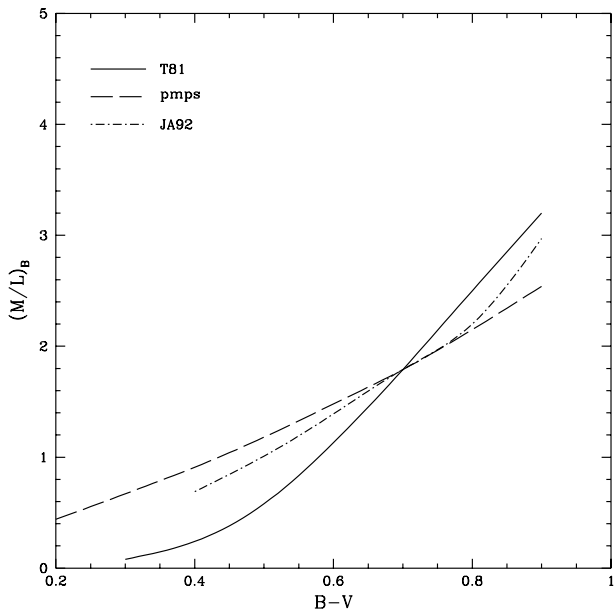,width=8.8cm}
\caption{
Total mass-to-light ratio (in B) versus B-V colour according to
Tinsley (T81), the described poor man's population synthesis (pmps), and
according to Jablonka \& Arimoto (JA92). The curves have been scaled
to coincide at $(M/L)_B = 1.79$ for B-V = 0.7. Despite its simplicity,
the predicted mass-to-light ratio for the pmps is similar to that of the
sophisticated population synthesis models.
}
\end{figure}

The present paper deals with the amount of mass in a galactic disc,
which is given by the observed velocity dispersions. Therefore it is
possible to fix the absolute scale of the mass-to-light ratio when 
comparing dispersions with luminosities and colours.
To that aim a few relations will be derived for an exponential
disc, leading to a relation which can eventually be compared with
the observed dispersions in Fig. 1. For an exponential old
disc
\begin{equation}
L_{\rm od} = 2\pi ({\mu}_0)_{\rm od} h^2,
\end{equation}
\begin{equation}
{\sigma}_0 = ({\mu}_0)_{\rm od} \; \left( {{M}\over{L}} 
\right)_{\rm od},
\end{equation}
and
\begin{equation}
v_{\rm max,od} = \vmaxd = 0.88\; \sqrt{\pi G {\sigma}_0 h},
\end{equation}
(Freeman, 1970) such that
\begin{equation}
v^4_{\rm max} = 0.3 \pi G^2 ({\mu}_0)_{\rm od}
\left( {{M}\over{L}} \right)^2_{\rm od} L_{\rm od},
\end{equation}
where $L_{\rm od}$ is the total luminosity of the old disc,
$({\mu}_0)_{\rm od}$ the central surface brightness of the old disc
in linear units e.g. $L_{\odot} {\rm pc}^{-2}$,
${\sigma}_0$ the central surface density and $h$ the scalelength.
Equation (9) holds exactly and as noted above, $(M/L)_{\rm od}$
is considered a universal constant.
\section{All discs have the same colour, B-V = 0.7, and obey Freeman's law}
For such a situation Eq. (9) can be written as 
\begin{equation}
M_{\rm od} = -10\; {\rm log}_{10}(\vmaxd) + P,
\end{equation}
which is a kind of Tully-Fisher relation. It appears that
for an exponential old disc this TF relation holds with
a coefficient of exactly 10. Or,
\begin{equation}
\vmaxd = 10^{0.1P}\cdot 10^{-0.1M_{\rm od}}.
\end{equation}
For an exponential disc the maximum rotation (at 2.2$h$) is
related to the vertical velocity dispersion through Eq. (1),
which, when combined with Eq. (11) gives
\begin{equation}
\dispzn = A^{-1} \sqrt{ {{z_0}\over {h}} } \; 10^{-0.1M_{\rm od}}.
\end{equation}
This relation is equal to Eq. (19) in B93 for
\begin{equation}
A = 0.88 \; 10^{-0.1P}.
\end{equation}
For Eq. (12) a fit can be made to the observed velocity 
dispersions as a function of $M_{\rm od}$ by chosing a certain $h/z_0$
behaviour. In B93 three choices of this behaviour are presented: one where
$h/z_0$ is constant at five, secondly a functionality such that the
dispersion versus luminosity relation (Fig. 1) becomes linear, and thirdly,
an intermediate situation where $h/z_0 = 0.6M_{\rm od} + 17.5$.
All three give a satisfactory fit to the dispersion data yielding a
disc TF relation (Eq. 10) with almost the same constant $P$. Still, the last
behaviour is preferred.
This is because then the $h/z_0$ value is somewhat larger
for the smaller galaxies as might have been observed (Bottinelli et al 1983;
Heidmann et al. 1972). In addition, the fit to the observed
dispersions is marginally better than for the $h/z_0 = 5$ 
(van der Kruit \& Searle 1981a, b, 1982) case (see Fig. 8 in B93).
A linear dispersion versus $M_{\rm od}$ relation
leads to the undesired property that for the least massive galaxies
the velocity dispersion in the disc becomes negative.

Observations of $h/z_0$ values are scarce and it is not a priori
predictable if and how $h/z_0$ is related to galaxy size.
Therefore, at the moment, with the limited information available,
the best suited linear functionality is adopted. Individual deviations
in $h/z_0$ values will certainly be the largest source of scatter
in any diagram of velocity dispersion versus galaxy size.
\begin{figure}[htbp]
\psfig{figure=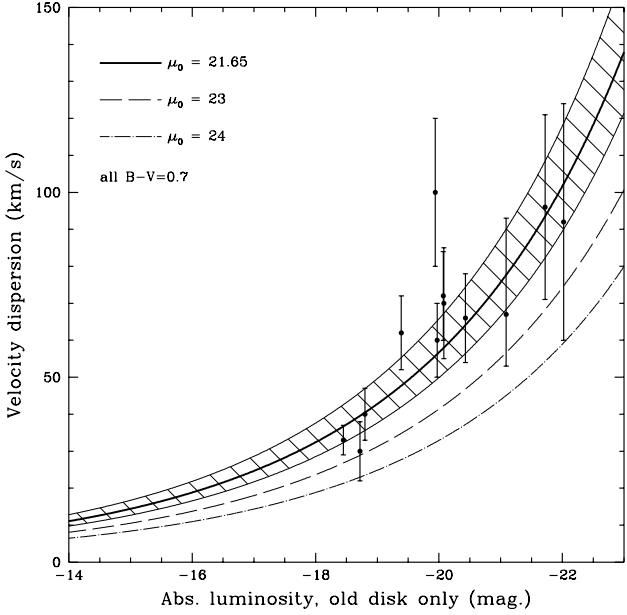,width=8.8cm}
\caption{
The observed velocity dispersion values of Fig. 1. Given by
the solid line and shaded area is a fit to these data
of Eq. (12) for the adopted $h/z_0$ behaviour. The data are
for galactic discs with on average ${\mu}_0 \sim 21.65$
B-mag. arcsec$^{-2}$ and B-V = 0.7.
Also indicated are the expected dispersion functionalities
for lower surface brightness galaxies with ${\mu}_0 = 23$ and 
24 B-mag. arcsec$^{-2}$.
}
\end{figure}

Taking $h/z_0 = 0.6 M_{\rm od} + 17.5$, Eq. (12) can now be compared
with the observed dispersions. This has been done in Fig. 8b in 
B93 and presently in Fig. 4. The best fit is achieved for
$A = 0.75$. By eye an error estimate has been made which is shown in
Fig. 4 as the shaded area around
the best fit for $A = 0.75 \pm 0.1$. From this a $(M/L)_B$ ratio
was derived for B-V equal to 0.7 of 1.79 $\pm$ 0.48. Substituting
the value found for $A$ into Eq. (13) and Eq. (11) one gets
\begin{equation}
\vmaxd = (1.17 \pm 0.16)\; 10^{-0.1M_{\rm od}},
\end{equation}
providing the maximum rotation of a single colour (B-V = 0.7) galactic disc
where $M_{\rm od}$ is given by Eq. (2) as $M_{\rm od} = M_{\rm tot} + 0.5$.
In Fig. 5 this maximum rotation of the disc only is compared
with the observed maximum rotation for
the total galaxy. The observations are for optical emission line
rotation curves of a sample of Sc and Sb galaxies
by Rubin et al. (1985) and of the galaxies by Mathewson et al. (1992).
For the latter absolute B magnitudes were obtained from the ESO-LV
catalogue (Lauberts \& Valentijn 1989) by Rhee (1996). 
The shaded area shown in Fig. 5
corresponds to the error given by the
shading in Fig. 4. It is obvious, comparing the disc-only
TF relation (Eq. 14) with the data in Fig. 5 that the
maximum rotation of the disc is considerably lower than
the observed maximum rotation. In fact, this is the
63\% criterion cast into a Tully-Fisher representation.
This now provides the possibility
to investigate the consequences for
a galactic disc with less restricted parameters.
\section{Discs with different colours and different central
surface brightnesses}
To investigate this general case Eq. (9) is rewritten to
\begin{equation}
v^4_{\rm max} = 0.3 \pi G^2 {\mu}_{0,F} 
{{({\mu}_0)_{\rm od} }\over{{\mu}_{0,F} }} \left( {{M}\over{L}}
\right)^2_{\rm od} L^{\rm od},
\end{equation}
where ${\mu}_{0,F}$ is Freeman's value which is constant
(by definition).
Use
\begin{equation}
{{ {\mu}_0^{\rm od} }\over{ {\mu}_0 }} = 
 {{ L_{\rm od} } \over{ L_{\rm tot} }} = 10^{0.4ct},
\end{equation}
and one finds
\begin{equation}
\vmaxd = {\rm const} \ast \left( {{ {\mu}_0 }
\over{ {\mu}_{0,F} }} \right)^{1/4} \;
10^{0.1ct}\;
 10^{-0.1M_{\rm od} }.
\end{equation}
Define the average colour correction term 
${\langle}ct{\rangle}$
for the sample
for which dispersions have been measured.
The sample has
galaxies with colours all close to B-V = 0.7 so that $\ct
 \sim -0.5$.
Insert this and Eq. (1) into Eq. (17) to get
\begin{eqnarray}
{\langle}v^2_z{\rangle}^{1/2}_{R=0} &=& 
{{\rm const}\over{0.88}} 10^{0.1\ct} \sqrt{ {{z_0}\over{h}}}
{{ 10^{0.1ct} }\over{ 10^{0.1\ct} }}\nonumber \\ 
 & & \cdot \left( {{ {\mu}_0 }\over{ {\mu}_{0,F} } } \right)^{1/4}
\; 10^{-0.1M_{\rm od} },
\end{eqnarray}
which can in principle again be fitted to Bottema's sample
of galactic disc dispersion measurements.
For these $ {\mu}_0 \sim {\mu}_{0,F}$ and $ct \sim \ct = -0.5$.
Adopt again $h/z_0 = 0.6M_{\rm od} + 17.5$, and one finds
for the same fit to the same data that:
\begin{equation}
{\rm const} \ast 10^{0.1\ct} = 1.17 .
\end{equation}
Substitute back into Eq. (17) to find:
\begin{equation}
\vmaxd = 1.17\; \left( {{ {\mu}_0 }\over { {\mu}_{0,F} } }
\right)^{1/4} 10^{0.1(ct-\ct)}\; 10^{-0.1M_{\rm od} }.
\end{equation}
Once more $M_{\rm od}$ can be converted to observed absolute magnitudes
\begin{equation}
\vmaxd = 1.17\; \left(  {{ {\mu}_0 }\over { {\mu}_{0,F} }} \right)^{1/4}
10^{-0.1M^B_{\rm tot}} \cdot 10^{0.2(ct + 0.25)},
\end{equation}
which is the most general expression for the maximum rotational
velocity of a disc and hence the principal result
of this paper.

The stellar velocity dispersion of a disc is found by substituting
Eq. (19) into Eq. (18) to get
\begin{eqnarray}
{\langle}v^2_R{\rangle}^{1/2}_{R=0} &=& 
1.33 \sqrt{ {{z_0}\over{h}} }
\left( {{  {\mu}_0 } \over { {\mu}_{0,F} }} \right)^{1/4}
10^{0.1(ct - \ct)}\nonumber \\ 
 & & \cdot \; 10^{-0.1M_{\rm od}}.
\end{eqnarray}
This relation is shown in Fig. 4 for B-V = 0.7 and 
${\mu}_0$ = 23 and 24 mag. arcsec$^{-2}$.
Different B-V values have not been plotted,
to avoid confusion, but the result of any preferred colour - surface
brightness combination can be inferred from Eq. (22).
Figure 4 shows that the LSB discs have lower stellar velocity dispersions
than normal discs with the same luminosity. 
However, this is only valid for an isolated stellar disc.
For small and/or LSB discs, for example, there may be large quantities of gas
available, which will increase the dispersion. Also a dark halo will
increase the stellar velocity dispersion in the outer parts of the
disc (B93). The extrapolation to $M_{\rm od} > -18$ in Fig. 4 is for the
adopted behaviour of $h/z_0$ as a function of luminosity. For a
different behaviour the result will, of course, be different.
\setlength{\unitlength}{1cm}
\begin{figure*}
\begin{minipage}{12.1cm}
\begin{picture}(12.1,12.4)
\psfig{figure=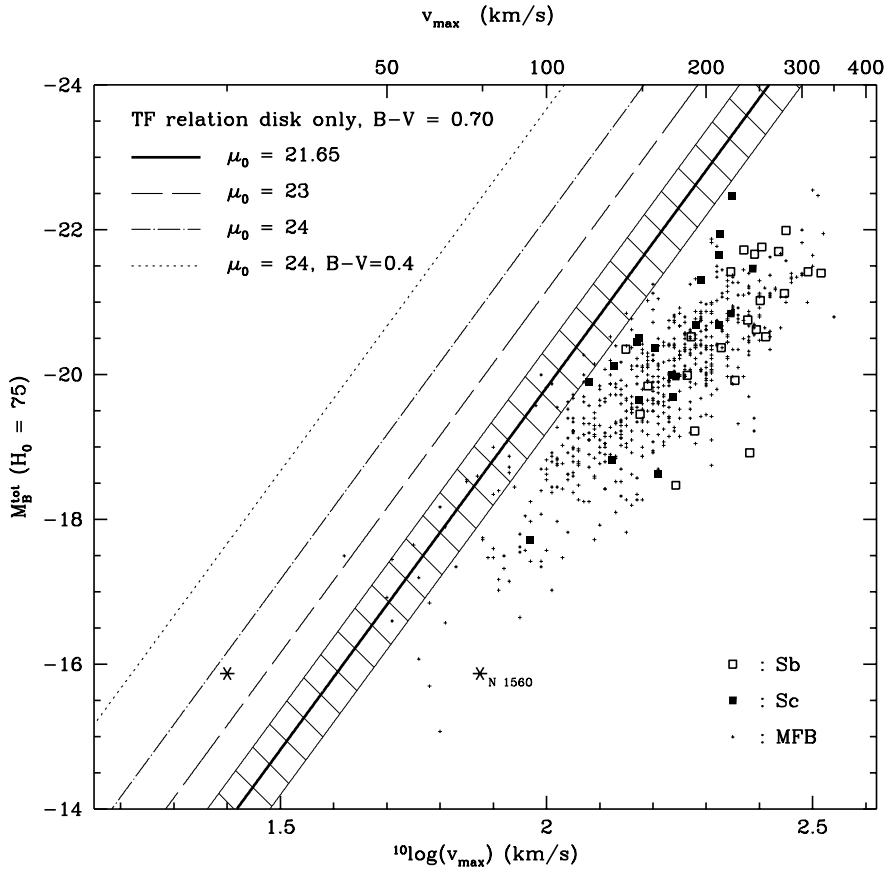,width=12.1cm}
\end{picture}
\end{minipage}
\hfill
\begin{minipage}{5cm}
\begin{picture}(5,0.1)
\end{picture}
\caption{
Representation of the TF relation for \underbar{observed} maximum rotations
given by the squares and crosses, and for the \underbar{disc only}
maximum rotation, as implied by the measured disc velocity
dispersions, given by the lines.
Observational rotations are derived from optical emission
lines for Sb and Sc galaxies  by Rubin et al. (1985) and for a sample of late
type galaxies by Mathewson et al. (1992, MFB). The disc-only maximum
rotation follows from Eq. (21) and is shown for a few central surface
brightness and colour combinations; the shaded area corresponds to the 
shading in Fig. 4.
The disc-only has considerably lower rotational velocities than observed
for the total galaxy, a difference becoming even more pronounced for the
lower surface brightness systems. As an example, for NGC 1560, the disc-only 
and total rotation is indicated by the left
and right asterisk respectively.
}
\end{minipage}
\end{figure*}

At this stage the implications of the general disc-only
TF relation (Eq. 21), will be investigated.
For ${\mu}_0 = {\mu}_{0,F}$ and B-V = 0.7,
Eq. (14) is retrieved as a special case which is already
depicted in Fig. 5 where disc-only rotations are compared
with total observed rotations of a galaxy.
In addition in Fig. 5
Eq. (21) is plotted for B-V = 0.7 and ${\mu}_0$ = 23 and 24
mag. arcsec$^{-2}$ and for B-V = 0.4 with ${\mu}_0$ is 24;
the regime of low surface brightness galaxies. For such objects
the disc-only rotation is at lower velocities than that of the 
normal surface brightness discs and at even lower velocities than
given by the observations. Since LSB galaxies seem to follow the same 
observed TF relation as non LSB galaxies (Zwaan et al. 1995) it
can thus be concluded that LSB discs contain a larger fraction
of dark matter than normal discs.
\section{The mass-to-light ratio}
For an exponential disc:
\begin{equation}
\dispzn = \sqrt{ \pi G {\mu}_0 \left( {{M}\over{L}}\right)_{\rm tot}
z_0 }.
\end{equation}
If we substitute Eq. (22) into Eq. (23) and
eliminate the dispersion one gets after some
algebra and unit conversions:
\begin{equation}
\left( {{M}\over{L}} \right)_B = 28.1 \;\; 10^{-0.2 M_B^{\odot} } \;
10^{0.4 ct} 10^{-0.2\ct}.
\end{equation}
Where $M_B^{\odot}$ is the absolute magnitude of
the sun. For $M_B^{\odot} = 5.48$ (Allen 1973) Eq. (24)
changes to
\begin{equation}
\begin{array}{lcl}
\left( {{M}\over{L}} \right)_B &=& 1.79\;\; 
10^{0.4ct} \; 10^{0.2(0.5 - \ct)}\\
&=& 1.93\;\; 10^{0.4(B-V)} - 1.88,
\end{array}
\end{equation}
%
being equal to the result already given in Eq. 5.
This shows that the observed dispersions actually fix the
mass-to-light ratio of the stellar population in an absolute sense.
There is a small discrepancy with B93 because in that paper
$M_B^{\odot} = 5.41$ was used (Allen 1963) leading to a coefficient
of 1.85 instead of 1.79 in Eq. (25). Therefore now for
$B-V = 0.7$, the value used for the one colour, one brightness
disc in B93, one finds $ct = -0.5$ and $\ct = -0.5 \Rightarrow 
(M/L)_B = 1.79$.
Note that the mass-to-light ratio of the general disc does \underbar{not}
depend on the central surface brightness of the disc. This is not surprising
since one of the assumptions was that the old disc population has 
the same mass-to-light ratio for all discs. Hence discs with equal 
colours also have equal total mass-to-light ratios irrespective of
the brightness.
\section{Two examples: NGC 6503 and NGC 1560}
To get a feeling for the implications of the results derived,
two specific examples will be discussed. For two galaxies for which
detailed and well resolved rotation curves have been measured, a decomposition
will be made of these curves into the contributions of the galactic
constituents. This is done for the maximum disc hypothesis situation and
for the disc contribution determined by velocity dispersions
and colour as given in Eq. (21).

The galaxy is supposed to consist of three components. First, a
disc with density distribution $\rho (z)$ as
\begin{equation}
\rho (z) = \rho (z=0)\; {\rm sech}^2 \left( {{z}\over{z_0}}\right),
\end{equation}
with thickness parameter $z_0$ being equal to 1/6 radial
scalelength. The radial density distribution was proportional to the
observed radial photometric profile. A rotation curve was calculated
according to Casertano (1983). Secondly, a thin gas layer
with surface density proportional to the observed radial \hi density
profile, multiplied with a factor 1.4 to account for Helium.
Thirdly, disc and gaslayer are embedded in a spherical pseudo
isothermal dark halo (Carignan \& Freeman 1985) with rotation
curve
\begin{equation}
v_{\rm halo}(R) = v_{\rm max}^{\rm halo}\;
\sqrt{ 1 - {{R_{\rm core}}\over{R}} {\rm arctan}\left(
{{R}\over{R_{\rm core}}}\right) }.
\end{equation}
A least-squares fit is made of the sum
of the individual contributions to the observed rotation
and best fitting parameters
$v^{\rm halo}_{\rm max}$ and $R_{\rm core}$ are determined.

The examples are NGC 6503 and NGC 1560; the first being a normal
surface brightness galaxy of moderate size and the second a typical example
of an LSB galaxy with ${\mu}^B_0$ = 23.23 mag. arcsec$^{-2}$.
A number of relevant galaxy parameters and results of
the fit are given in Table 1. Figures 6 and 7 show the
results for NGC 6503 and NGC 1560 respectively. For the
md hypothesis case and velocity dispersion implied case the fits are
equally valid. But there are appreciable differences in the
parameters of the individual components.
In the case where the disc mass is based on dispersions,
the disc is less massive and core radius and asymptotic halo velocity are
smaller than for the md hypothesis situation.
This implies that there is approximately a factor of two more
dark matter present in the disc region,
a conclusion which also applies for galaxies other than the two discussed
here.
\begin{figure}[htbp]
\psfig{figure=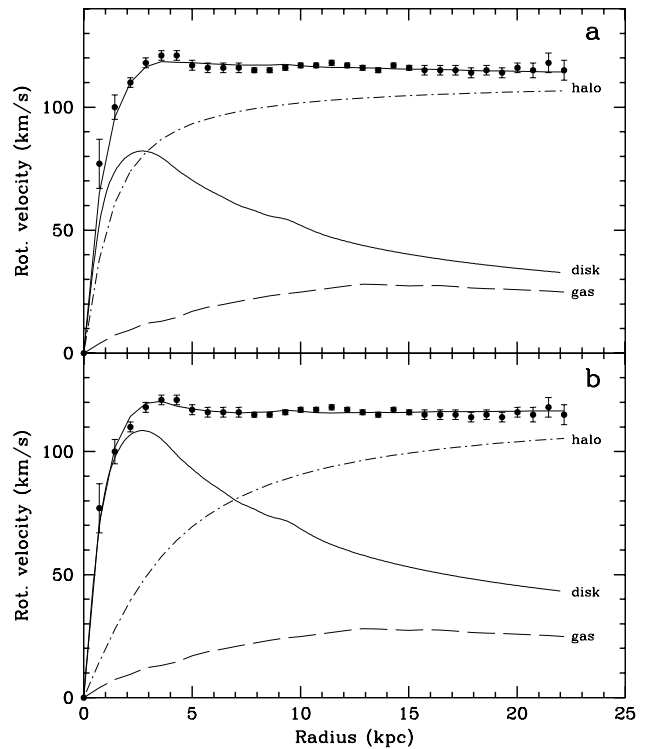,width=8.8cm}
\caption{
{\bf a and b}
Rotation curve decomposition for NGC 6503. The dots are the
observed values and upper solid line is the fit to these for:
{\bf a.} disc mass and rotation implied by stellar velocity dispersions
(Eq. 21) and {\bf b.} The maximum disc hypothesis.
}
\end{figure}

NGC 6503 offers the unique opportunity to compare maximum velocities of
a disc
derived in three different ways because it is in the sample for which
dispersions have been measured. Based on Eq. (21) a \vmaxds of 82
$\pm$ 11 \kmss is found. According to the 63 \% criterion ($\pm$ 10\%)
one finds \vmaxds = 76 $\pm$ 8 \kmss and the maximum velocity of the disc
can be calculated directly from the observed dispersions using
Eq. (1) with $h/z_0 = 6$ and $\disprh = \dispzn = 33 \pm 4$ \kmss such that
\vmaxds = 71 $\pm$ 9 \kms. The three results agree and show which
errors can be expected.

For NGC 1560, \vmaxds based on dispersion is 25 \kmss 
which is plotted in the TF relation in Fig. 5. The observed maximum
\hi rotation is 78 \kmss and optically one would obtain at $\sim$ 
4${{1}\over{2}}$ $h$ a maximum velocity of 72 \kms.
The average of these two is also given in Fig. 5 representing the
observed maximum rotation. The distance between disc only and observed
data point then gives a graphical
representation of the amount of dark matter. It illustrates nicely
the dominance of that component in LSB galaxies. 
\begin{table*}
\caption{Parameters of the galaxies}
\begin{flushleft}
\begin{tabular}{llll}
\noalign{\hrule\smallskip}
 & NGC 6503 & NGC 1560 & Ref.\\
\noalign{\smallskip\hrule\smallskip}
$(B-V)^o_{\rm T}$ & 0.57 & 0.57 & RC3/RC3, but corrected \\
Distance (Mpc) & 6.0 & 3.0 & Bottema (1989)/ Broeils (1992) \\
 $M_B^{o,i}$ (mag) & -18.76 & -15.87 & Sandage \& Tammann (1981)/
Broeils (1992) \\
${\mu}^B_0$ (mag. arcsec$^{-2}$) & 20.9 & 23.23 & Wevers et al. (1986)
+ Bottema (1989)/ Broeils (1992) \\
$ct$ & -0.78 & -0.78 & Eq. (3) \\
$v^{\rm disc}_{\rm max}$ (\kms) & 82 & 25 & Eq. (21) \\
$M_{\rm disc}$ ($10^9\; M_{\odot}$) & 5.42 & 0.52 & calculated \\
$R_{\rm core}^{\rm halo}$ (kpc) & 1.08 & 2.63 & lsq fit \\
$v^{\rm halo}_{\rm max}$ (\kms) & 111 & 89 & lsq fit \\
$v^{\rm disc}_{\rm max}$ md hyp. (\kms) & 108 & 43 & lsq fit \\
$M_{\rm disc}$ md hyp. ($10^9\; M_{\odot}$) & 9.46 & 1.53 & calculated \\
$R^{\rm halo}_{\rm core}$ md hyp. (kpc) & 3.37 & 15 & lsq fit \\
$v^{\rm halo}_{\rm max}$ md hyp. (\kms) & 119 & 243 & lsq fit \\
\noalign{\smallskip\hrule\medskip}
\end{tabular}
\end{flushleft}
\end{table*}
\section{Discussion and conclusions}
The developed description for maximum disc rotational velocities
(Eq. 21) can be applied for all discs, no matter the form
of the  observed rotation curve. Furthermore, an additional value of the
method is that it has a physical basis. Namely the equal
mass-to-light ratio for stellar populations having the same colour. 
This mass-to-light ratio is gauged by the observed velocity dispersion
of the sample of galactic discs. The present description is unlike
that of the md hypothesis, which is purely ad-hoc, or the
63\% criterion which is established observationally.
\begin{figure}[htbp]
\psfig{figure=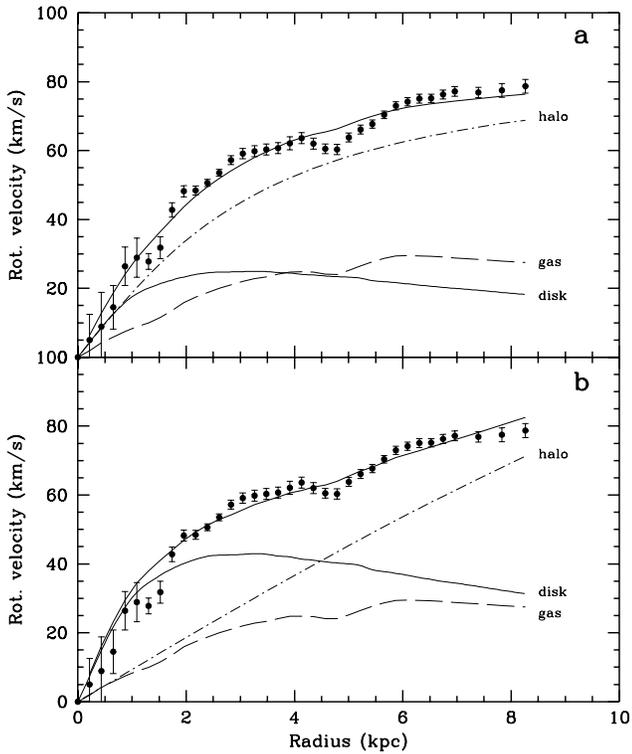,width=8.8cm}
\caption{
{\bf a and b}
As Fig. 6, but now for the LSB galaxy NGC 1560.
}
\end{figure}

However, there are some disadvantages of the method. To calculate
the maximum velocity of the disc according to Eq. (21), one needs
to know the absolute magnitude, central surface brightness, and
the colour of the disc. The absolute magnitude depends on the distance
and galactic and internal absorption correction. If these are ill
determined appreciable errors will result. To obtain the central
surface brightness a correction to face-on is needed, also a source
of errors. In addition the colour can never be determined with infinite
accuracy. Fortunately none of the parameters enters into Eq. (21)
in a dominant manner so that some error will not directly generate
a huge error in the maximum rotational velocity of the disc.

The poor man's population synthesis as described in section 2,
applies for a metallicity regime not too far from solar abundances.
For the majority of the nearby high surface brightness
(hereafter HSB; meaning ${\mu}_0 \sim {\mu}_{0,F}$) galactic discs
the metallicity can indeed be assumed to ly close to solar. 
However, for the LSB galactic discs abundance determinations
(Mc Gaugh 1994) indicate a metallicity content of typically 0.3 
to 0.1 times the solar values. This could pose a problem and therefore
the effect of a lower metallicity on the pmps will be  investigated.
Less metals in the same stellar population with the same age produce
a bluer B-V colour. For a solar abundance a B-V of 0.97
for the old disc population (5 to 10 Gyrs) was assumed in section 2.
For a metal poorer od population the colour can be determined from
Worthey (1994, his fig 34). Assuming a worst case scenario with
$Z = 0.1 Z_{\odot}$ the od B-V colour has to be decreased to $\sim$ 0.8.
The young disc colour is assumed to be roughly independent
of metallicity remaining at B-V = -0.03; indicated by
the observed small range in colours of young star
forming regions. The pmps was repeated for (B-V)$_{\rm yd}$ = -0.03 and
(B-V)$_{\rm od}$ = 0.8 giving a new colour correction term
$ct^{\prime} = 2.5\; {\rm log}_{10}[(1-0.973W)/1.116W]$.
Comparison with the solar abundance colour correction term $ct$
(Eq. 3) shows that $ct^{\prime} - ct = 0.30$ for all colours. This
means that for $Z = 0.1 Z_{\odot}$ the old disc population is 0.3
magnitudes brighter for the same observed colour; and hence the stellar
disc contains more mass generating a higher maximum rotation.
The latter can now be calculated simply from Eq. (21) when the metal
poor $ct^{\prime}$ is substituted resulting in a maximum disc velocity for
$Z = 0.1 Z_{\odot}$ being 15\% higher than that for a solar abundance.
Applied to the LSB example NGC 1560 this means that if
$Z$(N1560) $= 0.1 Z_{\sun}$ the maximum disc velocity should
be increased from 25 \kmss to 29 \kms. This still falls
substantially below the md hypothesis value of 43 \kms. It should
be kept in mind that this calculation is for a case
expected to be a limiting situation. The 15\% increase is
therefore a maximum.

In relation to LSBs there is another matter that has to be discussed.
The central surface brightness is an observed quantity; only a
correction to face-on is made but no absorption correction. HSB
discs are semi transparent with most of the extinction concentrated
in the inner regions (Huizinga \& van Albada 1992; Giovanelli 
et al. 1994). On the other hand, low luminosity spiral discs and 
LSB discs appear to be by and large transparent over the whole
extent of the disc (Bosma et al. 1992; Mc Gaugh 1994). Thus the
Freeman's law value derived for HSB discs of 21.65 B mag. arcsec$^{-2}$
is compromised by extinction with respect to LSB discs.
In order to correctly compare central surface densities of
HSB and LSB systems a correction should be applied to the
observed central surface brightness. Or, such a correction has
to make the surface brightness of HSBs larger, or, that of LSBs smaller.
Either way, in calculating maximum disc rotations the
central surface brightness quotient ${\mu}_0/{\mu}_{0,F}$ has to be
lowered for LSB systems with the typical amount of extinction in HSB discs.
Various studies indicate that this typical extinction 
ranges between 0.5 to 1 mag. in B (Keel 1983; Andredakis \& van der Kruit
1992; White \& Keel 1992; Knapen \& van der Kruit 1991; Byun et al. 1994;
Huizinga 1994). Inserting this correction in Eq. (21) results in a
lower maximum disc rotation for LSB systems of the order of 10 to 20\%.

According to the scheme developed in this paper a maximum disc
rotational velocity can be calculated for a galactic disc akin to
that of the solar neighbourhood. The lower metallicity
in LSBs would lead to a higher rotational velocity of at most 15\%.
On the other hand a different dust content of LSBs would lead to
a lower rotational velocity of 10 to 20\%. There is no direct
evidence that lower metals in discs is accompanied by less dust,
but it is more than likely. Therefore it is expected that in LSBs both
effects approximately cancel such that the original description also
applies for these systems. Moreover, effects of up to 15\% will leave
all conclusions essentially unchanged. 

The maximum disc velocity following from Eq. (21) is for a strict
exponential disc. In that sense it can be applied very well in a
statistical way comparing galaxies with one another as for example
in Fig. 5. For individual cases, when the photometric profile 
deviates strongly from exponential, the method can in principle
be extended to be used in  a radially differential way. 

Finally, a compilation of the main conclusions:
\begin{itemize}
\item
A general relation has been derived giving the maximum rotation of a
galactic disc as a function of its absolute magnitude, central
surface brightness, and colour. As a physical basis for this
relation serves an adopted universal M/L ratio as a function
of colour.
\item
This relation and the involved M/L ratios are fixed in
an absolute sense by the observed stellar velocity dispersions.
\item
Comparing derived maximum rotations of a disc with
the observed total rotation shows that even in the disc region,
normal galaxies contain an appreciable amount of dark matter.
Disc only rotations are lower by a factor of 0.7 compared
to the rotation implied by the maximum disc hypothesis.
\item
Low surface brightness galaxies contain an even larger
amount of dark matter.
\item
Derived disc M/L ratios are 1.79 $\pm$ 0.48 in the B-band
for a B-V of 0.7. This M/L value is around a factor
or two lower than the md hypothesis values.
\end{itemize}
\begin{acknowledgements}
I thank W. de Blok, R. Giovanelli, J. van der Hulst, 
and M.-H. Rhee for stimulating
discussions, encouragement and criticism. The Kapteyn Institute is
acknowledged for hospitality and support.
\end{acknowledgements}

\end{document}